\documentstyle[aps,pra,psfig,epsf,epsfig]{revtex}
\begin{document}
\draft
\twocolumn
\title{
Formation of Giant Quasibound Cold Diatoms by  Strong Atom-Cavity Coupling}
\author{B. Deb and G. Kurizki}
\address{
 Department of Chemical Physics, Weizmann Institute of Science,
76100 Rehovot, Israel} 
\date{\today}
\maketitle
\begin{abstract}
We show that giant quasi-bound diatomic complexes, whose size is typically
hundreds of nm, can be formed by intra-cavity cold diatom photoassociation or
photodissociation in  the strong atom-cavity coupling  regime. 
\end{abstract}
Cold atoms exchanging single photons with cavity fields have been predicted to
give rise to a variety of fascinating motional effects,  both in the
"good-cavity" (strong-coupling) regime of Rabi oscillations \cite{englert} and
in the "bad-cavity" regime of nearly exponential decay \cite{kuriz1}. Here we
pose a question that has not been considered thus far: what happens when two
identical cold atoms exchange a photon in the strong-coupling ("good-cavity")
regime during a collision or diatomic dissociation? We show here that this
regime gives rise to a novel, hitherto unexplored, interplay of molecular
dynamics and cavity QED effects. The resulting two-atom dynamics is drastically
modified at interatomic separations typically exceeding by 2 orders of
magnitude  those of currently investigated cold-atom collisional resonances in
magneto-optical traps \cite{inouye}. The predicted modification is due to the
possibility of {\it diatomic quasibinding} (scattering resonance) in a
potential well formed by the competing effects of the resonant dipole-dipole
(RDDI)  interaction and strong atom-cavity coupling. These  effects have
previously been  shown by our group to {\it suppress energy exchange} between
static atoms in a high-Q cavity \cite{kuriz}.  The regime considered here
drastically differs  from that of atoms coupled by  the RDDI in a "bad cavity"
\cite{meystre} or from that of coherently driven atoms in a cavity 
\cite{meyer}.  The quasibinding scattering potential formed by the RDDI and
strong atom-cavity coupling  is  shown to support many vibrational diatom
states. These states can be revealed in high-Q cavities by unusual spectral
structures in resonance fluorescence and Raman scattering from colliding or
dissociating pairs, as well as by sharp variation of the scattering
cross-section as a function of their center-of-mass kinetic energy. 

The model which we consider is that of a pair of cold two-level atoms
interacting with a single-mode cavity field. Cavity losses (damping) will be
accounted for later on. In the  center-of-mass (COM) frame, the radial part of
the interaction Hamiltonian $H$  can be split into two  parts.  The kinetic
part is 
\begin{equation}
H_{k} = - \frac{\hbar^{2}}{2 \mu} \frac{d^{2}}{dR^{2}} 
 - \frac{\hbar^{2} l(l+1)}{2\mu R^{2}}
\end{equation}
where $ \mu = \frac{m_{A} m_{B}}{m_{A} + m_{B}}$ is the reduced mass,  $ R=\mid
\vec{R}_{A}-\vec{R}_{B} \mid$ is the  separation between atoms A and B  and $l$
represents the angular momentum quantum number.  The other part is the
adiabatic Hamiltonian \cite{kuriz}
\begin{eqnarray} 
H_{ad} &=& H - H_{k} \nonumber \\
&=& \hbar\omega_{A} \sigma_{z}^{(A)} + \hbar \omega_{B} \sigma_{z}^{(B)} 
+ \hbar \omega_{c} a^{\dagger} a \nonumber\\
  & + &  \hbar  ( \kappa_{A } a \sigma_{+}^{(A)}
 +   \kappa_{B } a \sigma_{+}^{(B)} + h.c. )  \nonumber \\ & + &
\frac{\hbar C_{3}}{R^{3}} ( \sigma_{+}^{(A)} \sigma_{-}^{(B)} + 
\sigma_{-}^{(A)} \sigma_{+}^{(B)} )
\end{eqnarray}
Here $\omega_{c}$ is the frequency of the cavity-mode, $\omega_{A} \simeq
\omega_{B}$ are the atomic transition frequencies,  and $\kappa_{A,B }$ are
the  coupling parameters of the atom A(B) to the cavity mode; $C_{3}$ is the
dipole-dipole coupling coefficient, $\sigma_{\pm}$ and $\sigma_{z}$ are the
atomic pseudo-spin operators and $a(a^{\dagger})$ is annihilation(creation)
operator of the field mode. 

The adiabatic Hamiltonian in Eq.(2) is exactly solvable \cite{kuriz} in the
basis of  three coupled atom-field states  $ \mid e_{A}, g_{B},0 >$, $\mid
e_{B}, g_{A},0 >$ and $\mid g_{A}, g_{B},1 >$  where $  e_{j} ( g_{j}) $ 
represents the excited (ground) state of the $j$th atom  (j=A,B) while 0 and 1
are the number of photons in the cavity mode.  We can write $H_{ad}  \chi_{i} =
\hbar\omega_{i} \chi_{i}$ ($i=1,2,3$), with $\omega_{1}<\omega_{2}<\omega_{3}$.
As discussed in Ref.\cite{kuriz}, for $R \rightarrow 0$, $\omega_{3}
\rightarrow \omega_{s} = \omega_{A} + C_{3}/R^{3}$, $\omega_{2} \rightarrow
\omega_{c}$,  and   $\omega_{1} \rightarrow \omega_{a} =  \omega_{A} -
C_{3}/R^{3}$.   This means that the near-resonant cavity mode is then decoupled
from the symmetric and antisymmetric states $ \mid s,a \rangle = 2^{-1/2} (\mid
e_{A}, g_{B},0 > \pm \mid e_{B}, g_{A},0 >$). The symmetric-state  and
cavity-mode contributions become increasingly hybridized as $R$ increases
\cite{kuriz}:   $\omega_{1} \simeq \omega_{a}$,  $\omega_{2} \simeq \frac{1}{2}
(\omega_{s}+\omega_{c} - \Omega)$, $\omega_{3}\simeq \frac{1}{2}
(\omega_{s}+\omega_{c} + \Omega)$; with $\Omega=\sqrt{2
(\kappa_{A}+\kappa_{B})^{2}+(\omega_{s}-\omega_{c})^{2}}$. The potentials
$\omega_{1}$ and $\omega_{2}$ exhibit a pseudo-crossing at a relatively large
interatomic separation ($ \sim 1000 a_{0}$, where $a_{0}$ is the Bohr radius)
depending on the coupling strength of the atoms with the cavity field.   The 
potential  $\omega_{2}$ has a minimum at the pseudo-crossing point ($R_{c}$),
and so it can behave as a potential well supporting  quasi-bound states. We
illustrate these  long-range adiabatic potentials in Fig.1 for  a pair of  Cs
atoms with levels $6S_{1/2}$ and $6P_{3/2}$ ($ \omega_{A}=\omega_{B}=3.5172
\times 10^{14} Sec^{-1}$),  with atom-cavity parameters as in the experiment of
Ref.\cite{kimble} and realistic  dipole-dipole \cite{marin} coupling parameter.
The larger the atom-field coupling strength,  the deeper is the potential well
($\omega_{2}$)  with   the minimum point $R_{c}$ shifted towards  shorter
separations(Inset(a) to Fig.1).   For the  cavity parameters of 
Ref.\cite{kimble},  the depth of the binding potential should lie in the
radio-frequency regime.  An  increase of the atom-cavity coupling strength by 2
orders of magnitude   will enable the binding potential $\omega_{2}$ to
support  bound states separated by microwave frequencies.  Inset(b) to  Fig.1
shows the effect of cavity-atom detuning on the value of the potential
$\omega_{2}$ at $R_{c}$.  For both red and blue detunings  $\omega_{2}$ acts as
a binding potential, but increasing  red detuning leads to more attractive
$\omega_{2}$  and reduced $R_{c}$, while the opposite is true for blue
detuning. 

The potentials drawn in  Fig.1 must be correlated to the short-range
potentials  \cite{marin} in which the interatomic repulsion due to the overlap
of electron charge distributions is much larger than the atom-cavity coupling. 
For the Cs($6P_{3/2}$)-Cs($6P_{1/2})$ system, the short-range limit ($R \le 28
a_{0})$  of  potential $\omega_{1}$ should be correlated to the A
$^{1}\Sigma_{u}^{+}$ potential, which at long ranges  corresponds to the
$S-P_{3/2}$ asymptote \cite{marin}. We  interpolate the curve $\omega_{1}$ so
that it can merge with the model A $^{1}\Sigma_{u}^{+}$ potential at
separations comparable to the  equlibrium position at $R \simeq 10a_{0}$.  
Since the field-atom coupling strength $\kappa_{A (B)}$ varies sinusoidally 
along the standing-wave cavity mode, a pair of ultracold atoms  is more likely
to become quasi-bound by the binding potential $\omega_{2}$ in the vicinity of
an antinode than elsewhere. As we  show in Fig.1, the internuclear separation
$R$ for this quasi-bound state and a cavity mode of wavelength  $\lambda_{c}$
typically satisfies the  criterion $R<<\lambda_{c}$ (the pseudocrossing
separation $R_{c} \simeq $ 10\% of $\lambda_{c}$ for Cs-Cs). Thus a high-Q
cavity loaded with ultracold  atoms may give rise to periodic distribution  of
giant quasibound diatoms near the cavity mode antinodes. Within the range of
separations $R$ for which the adiabatic potential  is significant, the 
atom-field coupling  parameters $\kappa_{A}$  and $\kappa_{B}$ remain nearly
constant.

In order to elucidate the formation of  quasibound states of two atoms by  cold
atomic photoassociation \cite{klepp} or diatomic photodissociation inside a
high-Q cavity, we adopt multi-channel scattering theory \cite{newton}. Let the
eigenfunction of the Hamiltonian $ H = H_{k} + H_{ad}$  with angular momentum 
$l$ be given by $\psi_{l} = \Sigma_{i=1}^{3} (\phi_{i} \chi_{i})_{l}$. From the
Schroedinger equation $ H \psi_{l} = E \psi_{l} $, we obtain, after some
algebra,
\begin{eqnarray}
\frac{d^{2} \Phi}{dR^{2}} + 2 \tau^{(1)}\frac{d \Phi}{dR} = 
\frac{2 \mu}{\hbar^{2}}( U + \frac{l(l+1)}{R^{2}} - E ) \Phi
\end{eqnarray}
where $\Phi = (\phi_{1} \hspace{0.1cm} \phi_{2} \hspace{0.1cm} \phi_{3})^{T}$
is the solution vector and $U=\hbar$ diag.$(\omega_{1}, \omega_{2}, \omega_{3})$ is  the
diagonal eigenvalue matrix, the nonadiabatic-coupling matrix $\tau^{(1)}$ has the
elements $ \tau_{ij}^{(1)} = \langle \chi_{i} \mid \frac{d}{dR} \mid \chi_{j}
\rangle = -\tau_{ji}^{(1)}$ ($i,j=1,2,3$), and  the higher order terms 
$\tau_{ij}^{(2)} = \langle \chi_{i} \mid \frac{d^{2}}{dR^{2}} \mid \chi_{j}
\rangle  $ have been dropped, consistently with the low-energy regime.  For the
sake of convenience, Eq.(3) can be transformed into the following form 
\cite{newton}
\begin{equation}
(\frac{d^{2}}{dR^{2}} +  P^{2} -  W - \frac{l(l+1)}{R^{2}} )\tilde{\Phi}  = 0
\end{equation}
where $\tilde{\Phi} = T \Phi $, and  $W= \frac{2 \mu}{\hbar^{2}} [T U T^{-1}- 
\hbar $ diag.$ \{\omega_{1}(R \rightarrow \infty), \omega_{2}(R \rightarrow
\infty), \omega_{3} (R \rightarrow \infty ) \} ]  $;   T is a transformation
matrix  $ T = exp(-\int_{R_{\infty}}^{R} \tau^{(1)} dR) $,  with  the lower
limit of the  integration at an arbitrary large  separation denoted by
$R_{\infty}$ where all the potentials are flat;   P is a $3 \times 3$ diagonal
matrix whose diagonal elements are the $ P_{i} = \hbar^{-1} \sqrt{2 \mu \{ E -
\hbar \omega_{i}(R \rightarrow \infty)} \}$,  $i=1,2,3$. 

Equation(4) represents a compact form of the scattering equations of three
coupled channels, which are to be solved under the following boundary
conditions: $\tilde{\phi}_{m l}^{(i)} (R \rightarrow 0) \sim 0$ and   $
\tilde{\phi}_{m l}^{(i)} (R \rightarrow \infty) \sim P_{m}^{-1} Sin(P_{i} R -
\frac{l \pi}{2}) \delta_{mi} + P_{i}P_{m}^{-1} Cos(P_{m}R - \frac{l \pi}{2})
K_{mi}^{l}$, where the reaction matrix $K$ is related to the $S$-matrix (see
below) by   $K^{l}= i (1-S^{(l)})(1+S^{(l)})^{-1}$. Here  $i$  labels the
incident channel, along with the angular momentum  $l$. From now on we focus on
s-wave ($l=0$) scattering,  which is appreciable for ultra-cold atoms
\cite{weiner},  and omit the label $l$. The regular wavefunction which is a
solution of the multichannel scattering equations  can then be expressed in 
matrix   form \cite{newton} as
\begin{equation}
\tilde{\tilde{\Phi}}(R) = J + \int_{0}^{R} dR'
G(R,R') W(R') \tilde{\tilde{\Phi}}(R')
\end{equation}
where $\tilde{\tilde{\Phi}}$ is a $3 \times 3$ matrix with its $i$th column
being the regular part of the solution vector $\tilde{\Phi}$ for the $i$th
incident channel,  $J$ and $G$ are  $3 \times 3$ diagonal matrices with the
$n$th diagonal element being $J_{n}=Sin(P_{n}R)/P_{n}$ and 
\begin{eqnarray}
G_{n} &=& P_{n}^{-1} 
Sin(P_{n}R) Cos(P_{n} R') \nonumber \\ 
 &-& P_{n}^{-1} Cos(P_{n}R)Sin(P_{n} R')
\end{eqnarray}
These solutions must be supplemented by the analysis of the poles of the
S-matrix 
\begin{equation}
S=P^{\frac{1}{2}}F(-P)F^{-1}(P)P^{-\frac{1}{2}}
\end{equation}
 which is expressed in terms of the Jost function \cite{newton}
\begin{equation}
F(P) = 1 + \int_{0}^{\infty} dR exp(i P R) W \tilde{\tilde{\Phi}}
\end{equation}

In the absence of cavity dissipation, the bound states are given by the real
poles of the $S$ matrix whose energy $E$ is below the lowest threshold, i.e.,
below $\hbar \omega_{1}(R \rightarrow \infty),$ so that all the corresponding
channel momenta  $\hbar P_{i} (i=1,2,3)$ have positive imaginary values. The
scattering resonances are  given by the complex poles of the $S$ matrix,  with
energies $E$ above the lowest  threshold, the imaginary part of the complex energy 
being negative.   Resonances  should be observed as a signature of the
cavity-confined quasi-bound states.  In  Fig.2 we show the variation of
scattering cross-section ($\sigma_{11}$) against energy with only one (lowest)
open channel, the remaining channels being closed. This variation reveals the
cavity-induced resonances.  They have an effective width $\Gamma_{eff} =
\Gamma_{R} + \Gamma_{c}$,   where $\Gamma_{R}$  is the resonance
width in an ideal (lossless) cavity  and $\Gamma_{c}$ is  the rate of cavity
dissipation (cavity linewidth). For the case of Fig.1 our  estimate yields  
$\Gamma_{R} \sim 2$ MHz. If we take \cite{kimble} $\kappa_{A} \simeq 120$ MHz
and $\Gamma_{c} \sim 40$ MHz, the resonances associated with individual
vibrational states of $\omega_{2}$ cannot be resolved. However, for cold
Rydberg atoms \cite{mour} in a high-Q microwave cavity \cite{haroche}, these
resonances are shown to be resolvable due to the much larger ratio of
$\kappa_{A}/\Gamma_{c}$.

One may observe the predicted resonances, e.g., by populating loosely bound
states of the $Cs_{2}$ potential A $^{1}\Sigma_{u}^{+}$ \cite{marin} that are
below the dissociation threshold by an amount ranging from  0 to 100 MHz, via
photoexcitation of the ground molecular potential X $^{1}\Sigma_{g}^{+}$.
Nonadiabatic effects at separations of a few nm will then produce a mixure of 
$^{1}\Sigma_{u}^{+}$ and $^{3}\Pi_{u}$ states \cite{kuriz2}. The alternative is
photoassociation \cite{klepp}: As the two ground-state (S-state) cold atoms 
approach each other inside the cavity, they should be excited by a single
photon. By tuning the frequency of this photon, the two atoms can be raised to
the desired adiabatic potential $\omega_{1}$ or $\omega_{2}$,  their relative
velocity can be controlled and determine the scattering-channel  inputs in
Eq.(5).

In order to evaluate the characteristics of resonance fluorescence associated
with the discussed effect, we have approximated the potential $\omega_{2}$
given in Fig.1 by  the Morse model, for which   we can determine the
vibrational states supported by that potential. The vibrational frequency
$\omega_{2v}$ in the quasibinding potential $\omega_{2}$    depends on the
orientation of the molecular axis relative to the cavity axis, since the
adiabatic Hamiltonian given in Eq.(2) is a function of the angle   between the
atomic dipole and the cavity field (via $\kappa_{A,B}$). By applying the
Landau-Zenner-Stueckelberg (LZS) \cite{newton}  semiclassical treatment of
nonadiabatic coupling near the pseudocrossing of $\omega_{1}$ and $\omega_{2}$,
we have calculated the energy-shifts of each of those vibrational states. The
Franck-Condon spectrum of fluorescence  for the transition from $v$th level of
$\omega_{2}$ to the ground state ($gg$) of the diatom has  been calculated  for  an
exciting photon  polarized along the X-axis,  and an emitted photon along the Z
axis, which is the direction of the cavity field.  It can then be expressed as
\begin{eqnarray} 
 I_{2v \rightarrow gg}(\omega) &=& C  \int_{\theta} d\theta Sin(\theta) W_{\theta} 
\{ \frac{\Gamma_{eff}^{2} }
{(\omega -\omega_{2v}(\theta))^{2} + \Gamma_{eff}^{2}} \} \nonumber \\
&\times& 
\mid \int dR <\phi_{2 v} \chi_{2} \mid a(\sigma_{+}^{(A)} 
+  \sigma_{+}^{(B)}) \mid  \phi_{gg} >  \nonumber \\
&\mid& g_{A}, g_{B},1 > \mid^{2}     
\end{eqnarray} 
Here C is a constant, $\Gamma_{eff}$ is the effective linewidth as before,
$\omega_{2v}(\theta)$ is corrected for LZS pseudocrossing effects and
$W_{\theta}$ is a weight factor, both depending on the random angle ($\theta$)
between the Z-axis and the molecular axis:   $W_{\theta} = Sin^{2}(\theta) $
and $W_{\theta} = Cos^{2}(\theta) $  \cite{kuriz2} for the $\Sigma$ and $\Pi$
symmetry,  respectively. The wavefunctions $\phi_{2v}$ and $\phi_{gg}$
represent the $v$th vibrational state in the  $\omega_{2}$ potential and the
radial ground state, respectively. For a $\phi_{2v}$ state with $\Sigma$
symmetry, the dipole  selection rule forces $\phi_{gg}$ to have the  angular 
momentum $l=1$ in   the S-S asymptotic ground state potential (-1/$R^{6}$).  We
have $\phi_{gg} \simeq R j_{1}(\sqrt{2 \mu E_{gg}} R/\hbar)$ \cite{weiner},
where $j_{1}$ is the $l=1$ spherical Bessel function and $E_{gg}$ is the
ground-state (S-S) energy, which is assumed to be of  the order of one-photon
recoil energy for the transition $P_{3/2} \rightarrow S_{1/2}$. We  display the
spectrum [Eq.(9)] in Fig.3. 

Spectral peaks in the range of  the potential depth  $\omega_{2}$ are
indications of the cavity-confined quasi-bound diatomic states.  Our estimates
show that for the parameters of Fig.1 and Ref.\cite{kimble}, the
vibrational-line spacing is well below  the cavity linewidth ($\Gamma_{c} \sim
40 MHz$). In order to resolve these lines properly, the 
$\kappa_{A}/\Gamma_{c}$ ratio should be further increased, e.g., by using cold
Rydberg atoms in a high-Q cavity \cite{mour,haroche}  (see Fig.2b). Yet, even in the absence
of optimal resolution, the spectral signature of the quasibinding effect in
Fig.3 is unambiguous, especially for the $\Pi$ symmetry.

To conclude, we have demonstrated several unusual dynamical features obtainable
in the strong-coupling cavity regime of diatomic collisions or dissociation. 
These features are associated  with  vibrational quasibound states, which can
be excited  by intracavity photoassociation of colliding  cold atoms via single
photon absorption at a  frequency below the normal dissociation threshold. An
alternative is intracavity excitation of a cold dimer by  single-photon
absorption at a frequency just below the dissociation threshold.  The
prediction of cavity-induced giant quasi-bound cold diatomic complexes is the
first indication that  modifications of molecular collision dynamics by cavity
QED  may be used to 'engineer' novel metastable  states of matter, which will
live as long as the cavity holds the photon inside. 

This work has been supported by ISF, EU(TMR) and Minerva grants. We thank Dr.
B. Galanti for his help with numerical computations.

\vspace{0.5cm}

{\bf FIG.1} Adiabatic potentials (in MHz) as a function of interatomic
separation (in Bohr radii $a_{0}$)  for Cs atoms,   $\omega_{c}-\omega_{A} =
1.0 MHz$, $\kappa_{A} = 120 MHz $, at short separations and near the 
pseudocrossing (equilibrium) position $R_{c} \sim  2000a_{0}$, $\kappa_{B} =
0.8 \kappa_{A} $. Inset(a): Depth of $\omega_{2}$ (in MHz) and  position
$R_{c}$ (in Bohr radius) as a function of atom-field coupling strength
$\kappa_{A}$ (in MHz). Inset(b):   Minimum of $\omega_{2}$ (in MHz) at $R_{c}$ 
as a function of cavity-atom detuning (in MHz) for red
($\omega_{c}<\omega_{A}$) and blue ($\omega_{c}>\omega_{A}$) detuning.  

\vspace{0.2cm}

{\bf FIG.2} (a) S-wave scattering cross section $\sigma_{11}$ ( in $cm^{2}$) 
as a function of energy in MHz or  momentum $\hbar P_{1}$  (in $10^{-22}$ gm
cm/sec) for two Cs-atoms sharing an optical excitation in a cavity  without
losses ($\Gamma_{c}=0, \Gamma_{R}=2$ MHz) and with loss ($\Gamma_{c}=5$ MHz,
dashed) with  parameters  as  in Fig.1.   (b) Idem, for  Rydberg Cs-atoms
sharing  an excitation near $\omega_{A}=600 GHz$, $\kappa_{A}=150 KHz$,
$\kappa_{B}=0.99\kappa_{A}$,  and  $\omega_{c}-\omega_{A}=1.0 KHz$ for an ideal
cavity (solid line) and dissipative cavity with $\Gamma_{c}= 2 KHz$ (dashed
lines). 

\vspace{0.2cm}

{\bf FIG.3} Spectrum (arbitrary scale) of   $\mid \chi_{2} \phi_{2v} \rangle
\rightarrow  \mid g_{A} g_{B}, 1 \rangle $ transition   for $\Sigma$ and $\Pi$
symmetry of the quasibinding state with $\Gamma_{eff}= 8 MHz$ and randomly
oriented diatomic axis. The parameters are as in Fig.1.
 
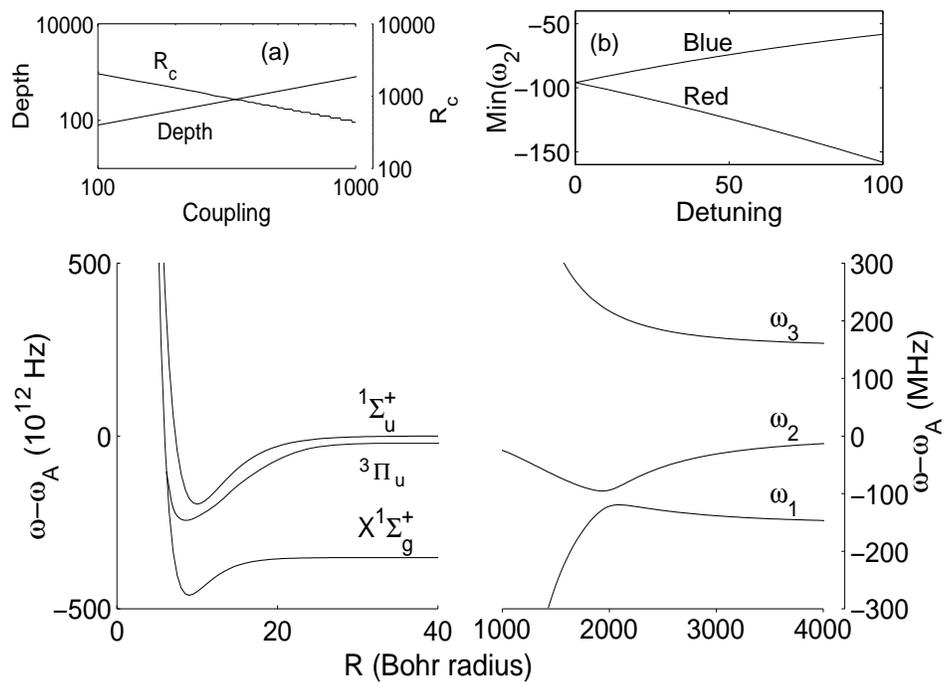
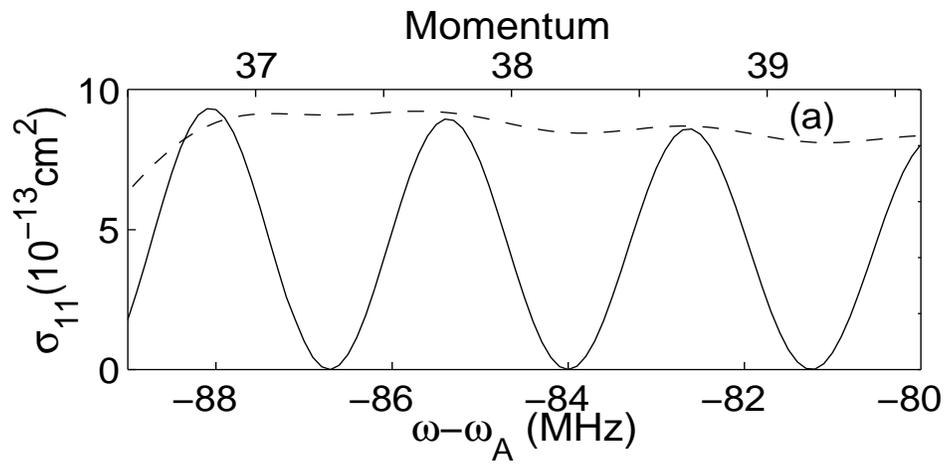
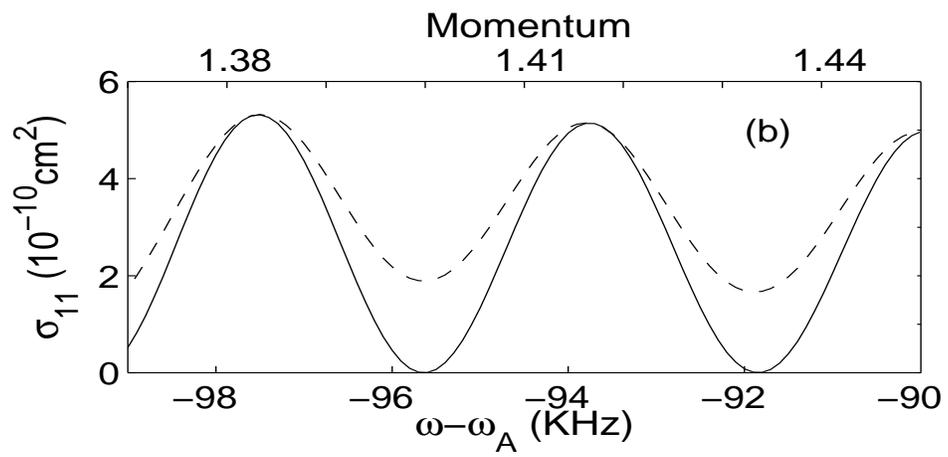
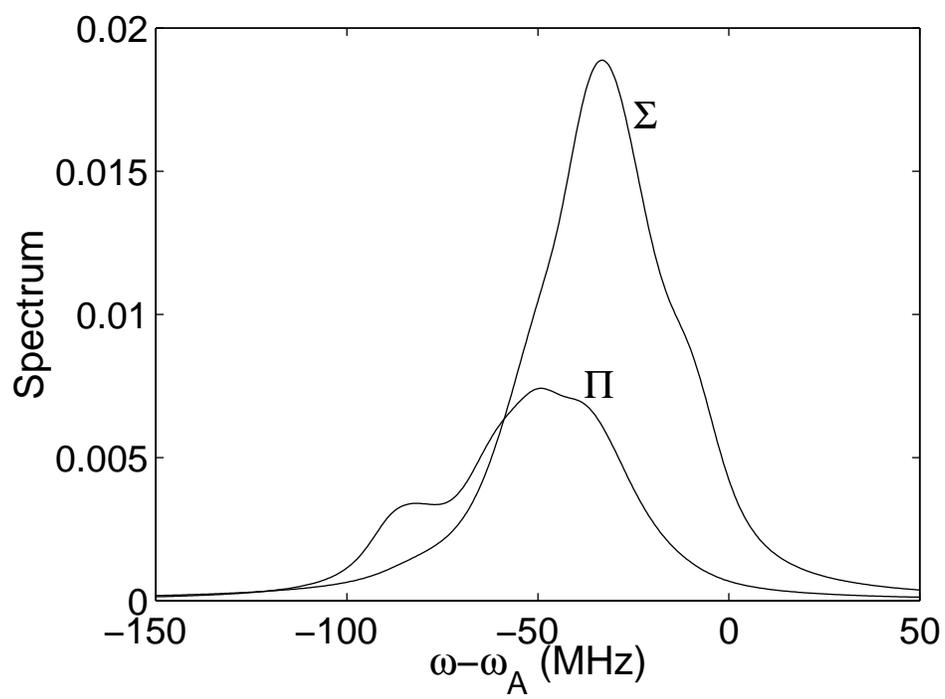
\end{document}